# Ultralong carrier lifetime of topological edge states in $\alpha$-Bi$_4$Br$_4$

*Pengcheng Mao[1,2], Junfeng Han,[1,2]\*, Jingchuan Zheng[1,2], Jinjian Zhou[3], Zhiming Yu[1,2], Wende Xiao[1,2], Dongyun Chen[1,2], Gang Wang[1,2], Jie Ma[1,2], Cheng-cheng Liu[1,2], Xiang Li[1,2], Qinsheng Wang[1,2], Junxi Duan[1,2], Hailong Chen[4,5]\*, Yuxiang Weng[4,5] and Yugui Yao[1,2]\**

[1]Key laboratory of advanced optoelectronic quantum architecture and measurement, ministry of education, School of Physics, Beijing Institute of Technology, Beijing, 100081, China.

[2]Micronano Center, Beijing Key Lab of Nanophotonics and Ultrafine Optoelectronic Systems, Beijing Institute of Technology, Beijing, 100081, China.

[3]Department of Applied Physics and Materials Science,

California Institute of Technology, Pasadena, California 91125, USA

[4]Beijing National Laboratory for Condensed Matter Physics and Institute of Physics, Chinese Academy of Sciences, Beijing, 100190, China.

[5]Songshan Lake Materials Laboratory, Dongguan, Guangdong, 523808, China.

[6]School of Physical Sciences, University of Chinese Academy of Sciences, Beijing, 100190, China.

**The rising of quantum spin Hall insulators (QSHI) in two-dimensional (2D) systems has been attracting significant interest in current research, for which the 1D helical edge states, a hallmark of QSHI, are widely expected to be a promising platform for next-generation optoelectronics. However, the dynamics of the 1D**

**edge states has not yet been experimentally addressed. Here, we report the observation of optical response of the topological helical edge states in α-$Bi_4Br_4$, using the infrared-pump infrared-probe microscopic spectroscopy. Remarkably, we observe that the carrier lifetime of the helical edge states reaches nanosecond-scale at room temperature, which is about 2 ∼ 3 orders longer than that of most 2D topological surface states and is even comparable with that of the well developed optoelectronics semiconductors used in modern industry. The ultralong carrier lifetime of the topological edge states may be attributed to their helical and 1D nature. Our findings not only provide an ideal material for further investigations of the carrier dynamics of 1D helical edge states but also pave the way for its application in optoelectronics.**

The field of topological insulator has undergone rapid development in the past decade and generated tremendous interest not only in finding out new topological phases, such as three-dimensional topological insulator [1,2], quantum spin Hall insulators (QSHI) [3,4], topological crystalline insulators [5-8], higher-order topological insulators [9-12], and the associated fundamental physical phenomenon, but also in searching for potential applications in electronics [13-15], spintronics [16-18] and optoelectronics [19-21]. Among them, experimental studies of the application of topological materials in optoelectronics is full of challengs, as it inevitably involves dynamic controls of nontrivial carriers. This challenge becomes more pronounced for one-dimensional (1D) states. Though those 1D states can already be probed by some sophisticated techniques like scanning tunneling microscopy (STM) [22-31], the optical response for 1D system is extremely weak and can not be directly detected by current technologies. For example, the state-of-the-art pump-probe techniques, such as time-resolved angle-resolved photoemission spectroscopy [32-34] and optical pump-probe spectroscopy [35-38], are powerful in detecting the carrier and spin dynamics of three-dimensional bulk and two-dimensional (2D) topological surface states of various topological insulators. However, they are not adequate for characterizing the dynamics of 1D topological edge states of 2D QSHI. This situation severely hinders both investigation and application of 1D topological edge states in optoelectronics.

To experimentally observe the dynamic process of 1D topological edge states of a 2D QSHI, a nature strategy is to stack 2D QSHI layers along the direction normal to the layer plane, as the existence of multiple 1D topological edge states can enhance the response signals. For detection with current technique, the number of the 1D edge states (e.g. the 2D QSHI layers) should be more than several dozens. Hence, such setup imposes several rigorous constraints on the 2D QSHI materials. First of all, the coupling between the 1D edge states localized at each QSHI layer should be very small to make such edge states are still of ('quasi-') 1D state. Second, a clean edge for each QSHI layer should be easily obtained. Although the 1D helical edge states are robust against disorders and defects due to topology nature, the optical response of the edge states, may be influenced by disorders and defects.

The family of bismuth halogenides, including α/β-$Bi_4Br_4$ and α/β-$Bi_4I_4$, has recently attracted great interest due to various fascinating properties, e.g. strong topological insulator [39], weak topological insulator [40,41], $C_2$-rotation protected topological crystalline insulators [42] and superconductivity [43-45]. α/β-Bi$_4$Br$_4$ and α/β-Bi$_4$I$_4$ are van der Waals (vdW) layer materials with each layer constructed by 1D atomic chains, making them quasi-1D materials. Recent studies predict that the vdW-type interlayer coupling of α-Bi$_4$Br$_4$ is rather weak and the monolayer Bi$_4$Br$_4$ is a QSHI with gapless topological edge states [41,46,47].

In this work, we observe the optical response of the topological edge states in stacked α-Bi$_4$Br$_4$ layers by using infrared-pump infrared-probe microscopic spectroscopy. Since both vdW-type interlayer and interchain coupling are weak in α-Bi$_4$Br$_4$, it is easy to obtain α-Bi$_4$Br$_4$ belts with (001)-oriented flat tops along the ab plane and straight and clean edges along the 1D chain axis (b-axis) via mechanical exfoliation, as shown in the images of optical microscopy (inset of Fig. 1(a)). As the coupling of the topological edge states localized different Bi$_4$Br$_4$ layers are very weak [41,46,47], the edge states of α-Bi$_4$Br$_4$ always maintain their 1D nature regardless of the number of the stacked layers. Thus, in α-Bi$_4$Br$_4$ belts, the signal of the optical response of 1D helical edge states can be extremely enhanced by 2 ∼ 3 orders (depending on the layer number of sample) without losing of the desired 1D nature. Based on the observed ultrafast pump-probe microspectroscopy, we can estimate the carrier lifetime of the edge states of α-Bi$_4$Br$_4$ belt. Surprisingly, we find the carrier lifetime is as long as ∼ 1 ns at room temperature, comparable with the typical carrier lifetime (∼ ns) of the optoelectronics semiconductors used in modern industry. For comparison, we notice that the carrier lifetime of 2D topological surface state of most reported 3D topological insulators is about several to tens of picoseconds [32-38,48-51], which is 2 ~ 3 order shorter than that of the edge state of α-Bi$_4$Br$_4$ belt. To our best knowledge, this is the first observation of optical dynamic of 1D topological helical edge states and revealing the ultralong carrier lifetime of the edge states. Such ultralong-lived carriers in the edge is benefit for optical control of the topological edge states and expected to be used to design novel optoelectronic devices.

The α-Bi$_4$Br$_4$ belts are obtained by mechanical exfoliation from the corresponding bulks [52]. We first measure the infrared absorption of an α-Bi$_4$Br$_4$ belt with a width of ~ 80 μm and a thickness of ~ 1.3 μm. Since the height of monolayer Bi$_4$Br$_4$ is about 1 nm, this belt contains more than one thousand of Bi$_4$Br$_4$ layers with 1D helical edge states, enabling us to detect the response of edge states using current technology. The infrared absorption spectra are collected from two typical regions of the belt [indicated in the inset of Fig. 1(a)]: center [blue curve in Fig. 1(a)] and side [red curve in Fig. 1(a)]. First, the optical gap for the bulk α-Bi$_4$Br$_4$ is about 0.2 eV (as indicated by the blue curve). The non-vanishing spectra for the energy smaller than 0.2 eV may be from the intraband transition. Second, the spectra collected at side contains the attributions from bulk states and edge states of α-Bi$_4$Br$_4$ and the substrate CaF$_2$. Due to finite diameter of light spot and negligible absorbance of substrate, the absorbance collected at side would be much lower than that collected at center. This is the case for energy higher than bulk band gap 0.2 eV, where the absorbance at side is much weaker than that at center. Surprisingly, one finds that for the energy lower than Eg, the absorption of side is stronger than that of the center, clearly showing the edge states have considerable contribution to the absorption spectra.

The carrier transfer dynamics of the α-Bi$_4$Br$_4$ belt was directly monitored with the infrared-pump infrared-probe microscopic spectroscopy (Fig. 1b, more experimental details in supplemental materials [52]). In these experiments, two kinds of incident femtosecond tunable mid-infrared pulse pump (0.5 eV and 0.17 eV) uniformly illustrate the belt to excite carriers in the bulk and side simultaneously. The evolution of the photo-excited carriers is then monitored by recording the absorbance change (ΔOD) in the mid-infrared region (0.15 ~ 0.26 eV) with a focused femtosecond probe pulse. The diameter of focal spot is ~20 μm at 7 μm wavelength. Figure 1 c & d show the temporal evolutions of ΔOD from 0.15 eV to 0.26 eV upon the 0.5 eV and 0.17 eV photo pump respectively with pulse energy of ~ 80 μJ/cm$^2$ from the belt side. A negative ΔOD can be clearly observed in the energy smaller than 0.2 eV (the bulk band gap), especially in the spectra under the 0.17 eV excitation. The negative signal (0.15 ~ 0.17 eV) is consistent with strong infrared absorption region of α-Bi$_4$Br$_4$ belt side (< 0.2 eV), indicating the negative signal may originate from the edge states. In addition, those negative signals are still observed even after 1000 ps relaxation [52], manifesting a kind of ultralong carrier lifetime.

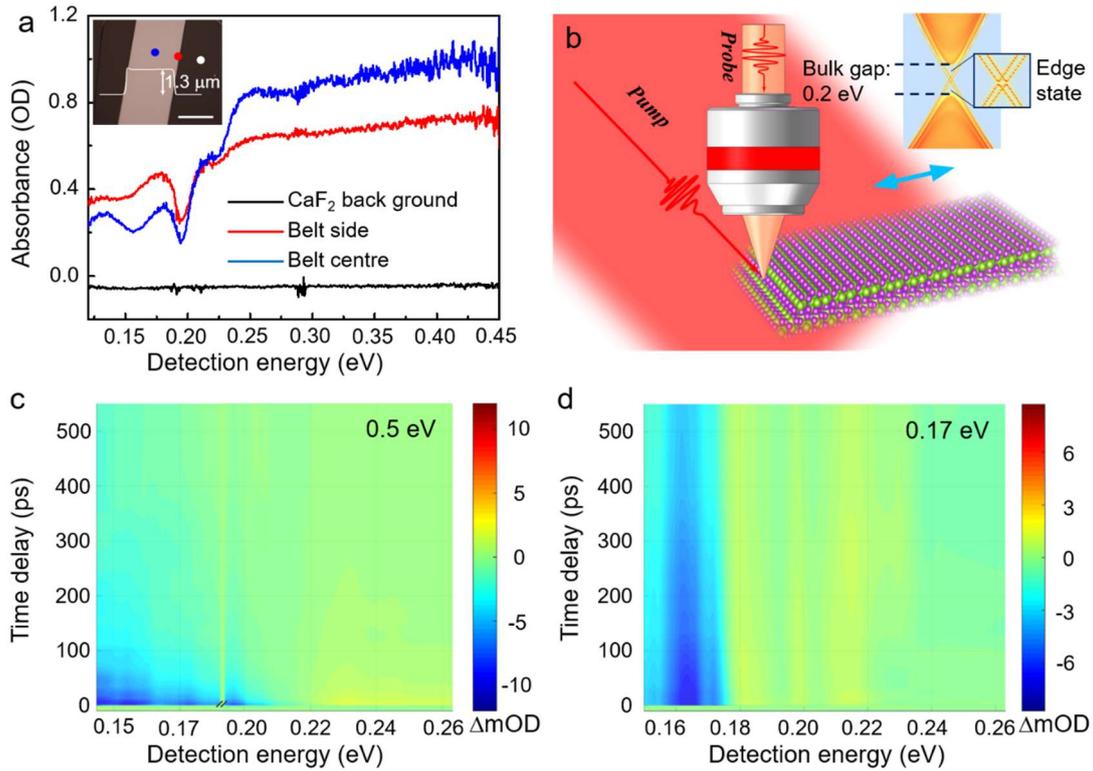

**Fig. 1 | Infrared absorption properties of α-Bi$_4$Br$_4$ belt. a,** Infrared absorption spectra acquired from the belt centre (blue curve), belt side (red curve) and the CaF$_2$ substrate (black curve). The insert is the OM image of the single-crystalline α-Bi$_4$Br$_4$ belt, showing a flat top and straight edges. Scale bar, 50 μm. The blue, red and white dots indicate the positions where the spectra are acquired. The infrared absorbance (OD), which is defined as OD = log(I$_0$/I), where I$_0$ and I are the intensities of the incident and transmission light, respectively. **b,** A schematic illustration of ultrafast infrared-pump infrared-probe microscopic spectroscopy microspectroscopy. **c, d,** Temporal evolutions of respective 0.5 eV and 0.17 eV excitation-induced absorbance change (ΔOD = OD$_{pump-on}$ - OD$_{pump-off}$) detected at 0.15 ~ 0.26 eV of the belt side. An obvious negative signal can be observed in the detected photon energy range of 0.15 ~ 0.17 eV.

To distinguish the carrier dynamic behaviors of bulk and edge states, we carried out the spatial resolved pump-probe analyses across the α-Bi$_4$Br$_4$ belt. The probe photon energy was fixed at 0.165 eV to avoid probing interband transitions between the bulk states. Figure 2 a & b exhibit the evolution of infrared ΔOD with detection areas from the belt center to the belt side under the incident pump light with 0.5 eV and 0.17 eV photon energies, respectively. With 0.5 eV excitation, an obvious positive ΔOD can be observed and relaxed to the ground state on a timescale of ~100 ps when the probe light passes through the belt center. By contrast, when the probe light pass through the belt side, the ΔOD signals changes to negative values for both 0.5 eV and 0.17 eV pump and relaxes to the ground state on a relatively longer timescale. In this experiment, the infrared probe photons with energy (0.165 eV) smaller than the bandgap can only be absorbed by the carriers in the edge states or the conduction band. Therefore, in the belt center (no edge states), more photoexcited carriers in the conduction band can enhance the absorption of 0.165 eV photons, contributing to the obvious positive signals. However, in the belt side, the decreased carrier density in the edge states reduces the infrared absorption after optical excitation and consequently leads to negative signals.

To get a qualitative understanding of these features, we drew the decay curves of ΔOD from both the belt center and the side, and performed a fitting analysis with the multiple-exponential functions. The best-fitted curves are shown as solid lines in Fig. 2c. In the belt center, the positive signal of ΔOD quickly reaches the maximum within 200 fs and then decays in ps timescale. The relaxation is decomposed into two decay components: a fast component ($\tau_1$) with characteristic time of 3 ~ 5 ps and a slow component ($\tau_2$) with characteristic time of 50 ~ 100 ps. Compared with the decay curve in the belt center, the signal in the belt side has an additional ultralong component with a characteristic time ($\tau_3$) longer than 1000 ps. Especially, with the pulse pump by 0.17 eV photons, the ultralong component becomes more significant (Fig. 2c). By comparing the amplitudes of those components in different relaxation processes (the inset of Fig. 2c), we can find that the ultralong component only appears in the belt side and its amplitude increases with smaller photon energy excitation. Furthermore, the ultralong carrier lifetime $\tau_3$ always exists in the different edges although with various local geometry and/or chemical bonds as shown in Fig. 2d, which suggests ultralong lifetime carrier behavior are related with the robust topological properties of $\alpha$-Bi$_4$Br$_4$. Such an ultralong lifetime is 2 ~ 3 orders longer than that of surface states in most three-dimensional topological insulators [32-38,48-51].

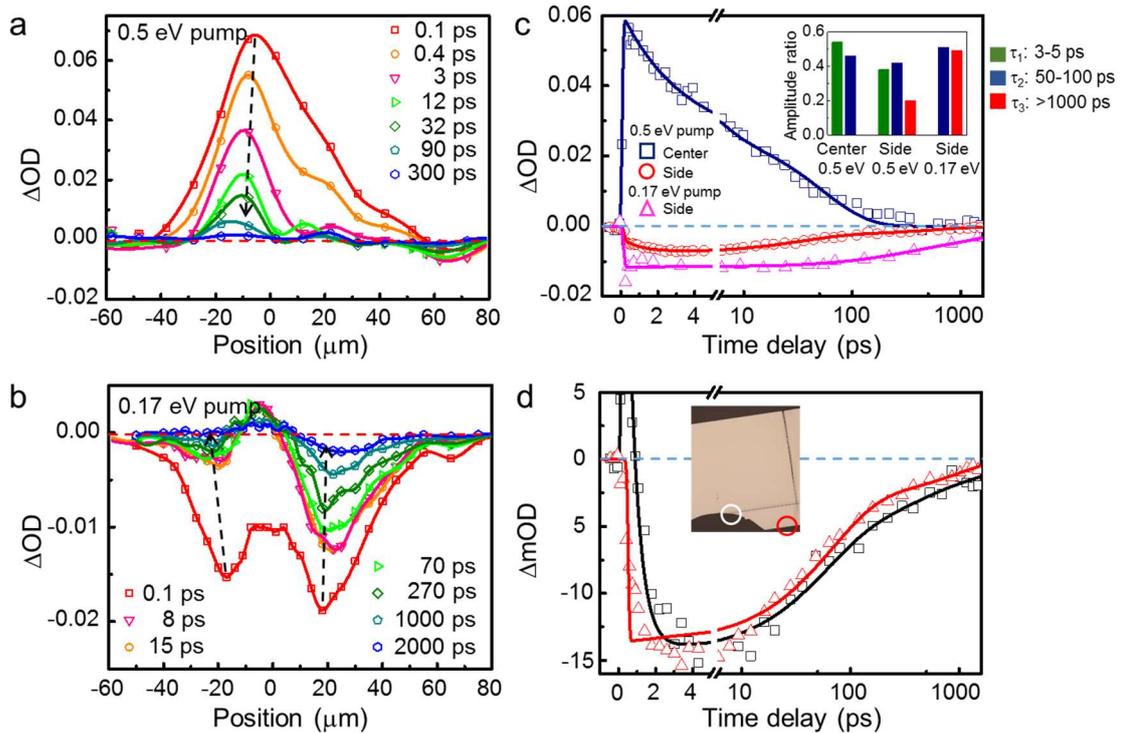

**Fig. 2 | Ultrafast pump-probe microspectroscopy measurements of the α-Bi$_4$Br$_4$ belt under different excited photon energies. a, b,** Temporal evolutions of the respective 0.5 eV and 0.17 eV excitation-induced ΔOD detected at 0.165 eV for a spatial resolved pump-probe analyses across the α-Bi$_4$Br$_4$ belt, where the belt center is located at the 0 μm position. **c,** Temporal evolutions of the 0.5 eV and 0.17 eV excitation-induced ΔOD detected at 0.165 eV of the α-Bi$_4$Br$_4$ belt center and side. The raw data (dots) are multi-exponentially fitted (curves). The insert shows the amplitude ratio of those relaxation components from different curves. **d,** Temporal evolutions of the 0.17 eV excitation-induced ΔOD detected at 0.165 eV of the α-Bi$_4$Br$_4$ belt side. The raw data (dots) are multi-exponentially fitted (curves). The red and white circles in the inset indicate the positions where the signals are acquired.

Considering less amount of the edge states compared with that of bulk states in the multilayer α-Bi$_4$Br$_4$ belt, the behavior of ΔOD with increased pump intensity in the belt side should be different to that in the belt center. For this sake, we repeated the measurement in the belt center and side with various pulse pump energies from 16 μJ/cm$^2$ to 690 μJ/cm$^2$ as shown in Fig. 3 a & b. First, we find that the magnitude of the signal in the belt center is proportional to the pump energy, while that in the belt side shows saturation effect. A simple saturation model can be used to describe the pump energy dependent signals [53]: $\Delta OD/OD \propto E/(E+E_s)$, where E is the pump energy and E$_s$ is the saturation energy. The fitted curve is shown in Fig. 3c, which indicates the saturation energy ~ 80 μJ/cm$^2$. According to our previous theory calculation [46,47], the density of edge state is estimated to be ~ 1 state/eV for each unit cell. Thus, the photons in the incident light with 80 μJ/cm² per pulse are enough to excite most carriers in edge states [52], making the ΔOD saturation. Second, we observe that the delay time has strong dependence on the pump intensities, decreasing with stronger intensities [see Fig. 3(d)]. The faster decay is caused by the enhanced electronelectron scattering or auger recombination with higher carrier density excited by higher pump energy [54].

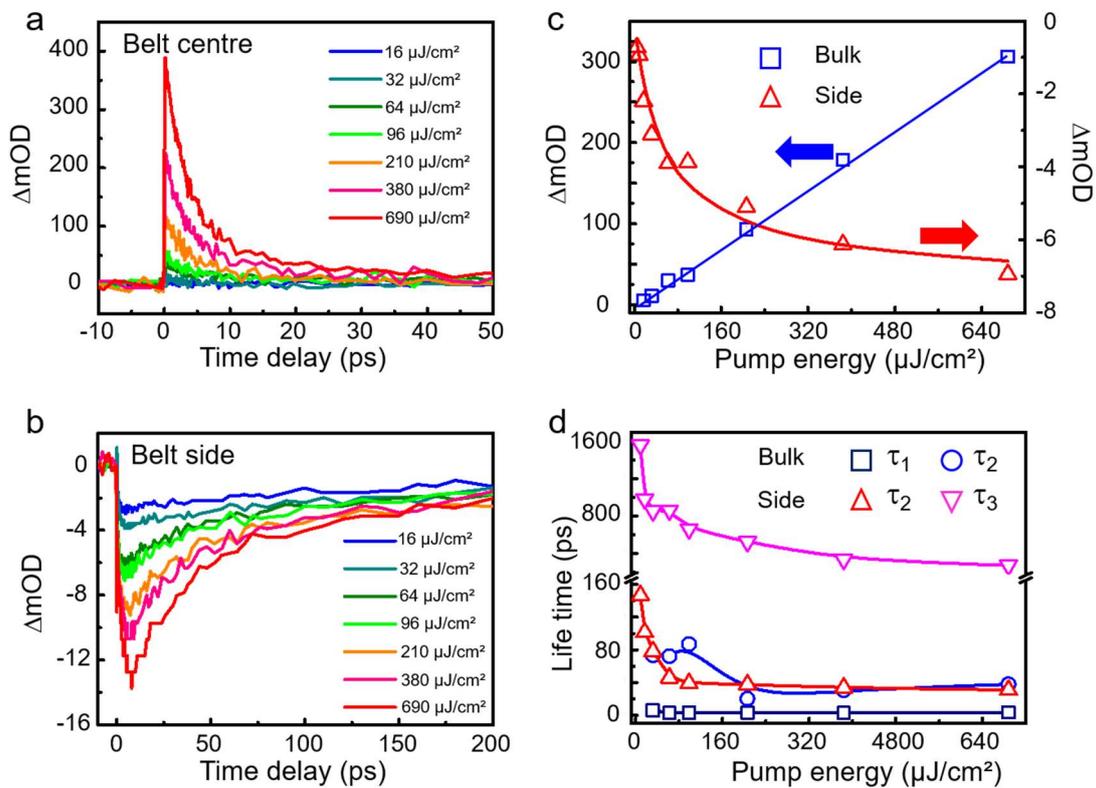

**Fig. 3 | Ultrafast infrared-pump infrared-probe microscopic spectroscopy microspectroscopy measurements of α-Bi₄Br₄ under different pump energies. a, b,** Temporal evolutions of the 0.5 eV excitation-induced ΔOD detected at 0.165 eV of the α-Bi₄Br₄ belt center and side under different pump energies. **c,** The magnitude of the ΔOD as a function of the pump energies. The blue solid line indicates a linear fit for the signals of the belt center, while the red solid line indicates a saturation

function fit for the signals of the belt side. **d,** The carrier lifetime as a function of the pump energies. The $\tau_1$, $\tau_2$, $\tau_3$ are three typical relaxation times observed in the α-Bi$_4$Br$_4$ as indicated in Fig. 2c.

The relaxation dynamics of photoexcited carriers are shown schematically in Fig. 4. At the belt center where the topological edge states are absent, the relaxation processes are similar to the typical carrier relaxation behaviors commonly observed in semiconductor materials (see Fig.4a) [55]. Process one ($\tau_1$: 3 ~ 5 ps) is the excited electrons relax to band edge and process two ($\tau_2$: 50 ~ 100 ps) is the relaxed electrons recombine with holes in the valence band (as illustrated Fig.4a). Intriguingly, at the belt side, besides the two relaxation times $\tau_{1(2)}$ observed in the bulk, another relaxation time $\tau_3 > 1000$ ps is observed. We contribute the ultralong relaxation time $\tau_3$ to the quasi-1D nature of the electronic states at the side, e.g. the 1D topological helical edge state of α-Bi$_4$Br$_4$. The reasons are the following. First, as discussed above, the (100) surface state of bulk α-Bi$_4$Br$_4$ shows the conduction band along (001) direction (normal to the plane) is almost dispersionless and the Fermi surface for electron doping is almost always open rather than close, indicating the interlayer coupling of the 1D topological states is vanishingly small. Thus, although the sample is more than one thousand layers thick, the states localized at the sample side are trivially stacked 1D topological states rather than a normal 2D surface state with closed Fermi surface (or Dirac cone). As a result, the edge states localized at different layers are weakly coupled or even decoupled if the edges of adjacent layers are not strictly aligned (see Fig. 4c) [47]. Hence, the direct recombination of electrons and holes at different edges and the inter-edge scattering are suppressed. Second, due to 1D helical nature, several possible scattering processes, which can strongly reduce the carrier lifetime, are suppressed in each monolayer (see Fig. 4b). For example, the backscattering caused by non-magnetic disorder and electron-phonon interaction is completely prohibited by the 1D helical nature and the direct recombination of electrons and holes is also suppressed due to spin flip involved. In addition, the electron-phonon scattering between edge states with the same spin polarization (see blue dashed line in Fig. 4b) is also a ultra-slow process due to the very limited scattering phase space (phonon energy < 20 meV, very few available final states within the 1D channel) [56,57]. Therefore, the helical nature of the 1D topological edge states combined with the weak coupling of these edge states lead to their ultralong carrier lifetime.

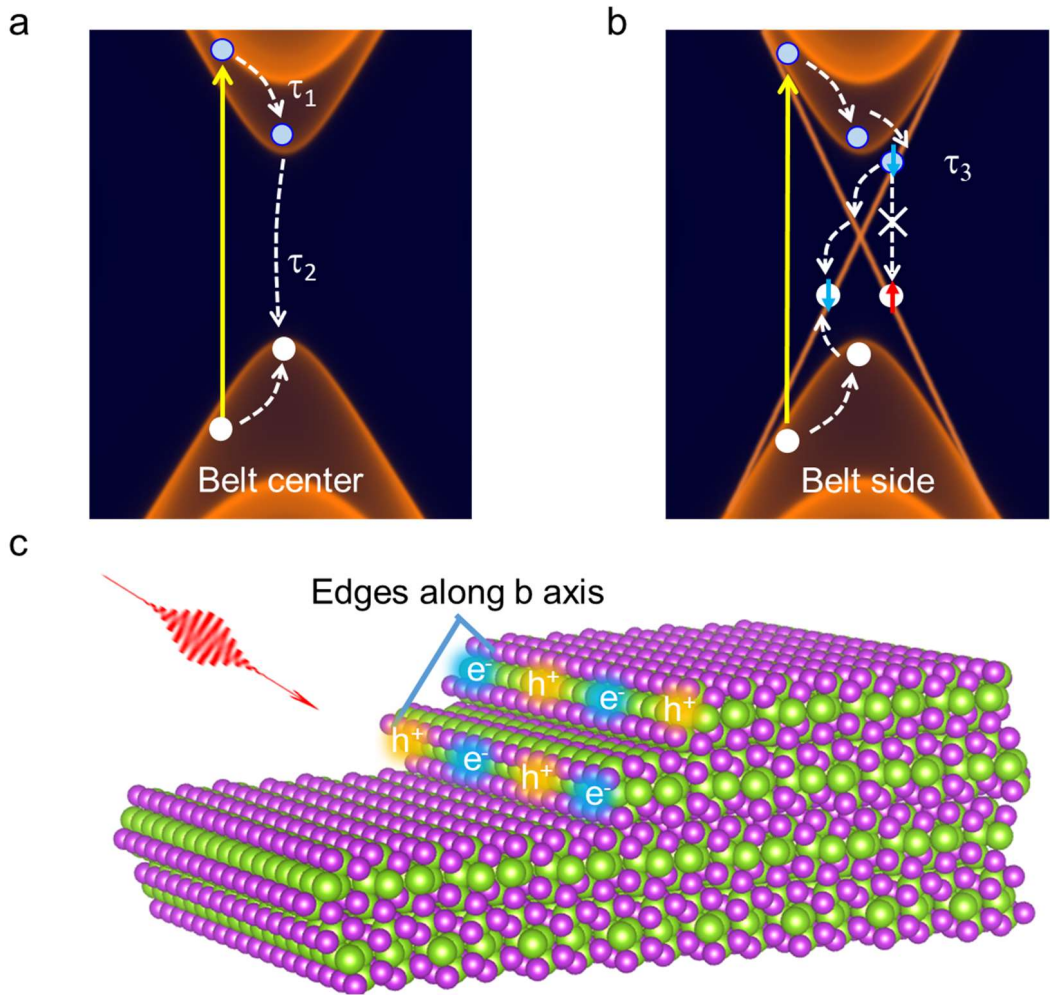

**Fig. 4 | Illustration of carrier dynamics of the edge states of α-Bi$_4$Br$_4$. a,** Schematic representation of the ultrafast excitation and subsequent relaxation processes of photoexcited carriers in the conduction band and the valence band. (1) Photoexcitation of electron-hole pairs; (2) hot carrier cooling due to electron-electron scattering or electron-phonon scattering; (3) recombination of electrons and holes in conduction band and valence band, respectively; **b,** Schematic representation of relaxation processes of photoexcited carriers in the edge states. (1) Direct recombination of electrons and holes is prohibited due to their opposite spin polarization; (2) the electron-phonon scattering with the same spin polarization is suppressed by the very limited scattering phase space (blue dashed line); **c,** Illustration

of the inter-layer scattering of carriers localized at different layers. The edges of different layers are not strictly aligned, forming step edges with decoupled states, which could significantly suppress the recombination of electrons and holes by the inter-layer scattering.

In summary, we use infrared-pump infrared-probe microscopic spectroscopy to investigate the carrier dynamics of $\alpha$-Bi$_4$Br$_4$. Except two trivial relaxation process with characteristic time of 3 ~ 5 ps and 50 ~ 100 ps, we observed a ns-scale relaxation process originated from the edge states of $\alpha$-Bi$_4$Br$_4$, 2 ~ 3 orders longer than the ps-scale lifetime of surface states in most three-dimensional topological insulators. With both experimental and theoretical analysis, we predict that the carriers with ultralong lifetime are from the 1D topological edge states of $\alpha$-Bi$_4$Br$_4$. This is first time for experimentally observing the optical response of 1D topological edge state. Our work not only reveals the ultralong carrier lifetime of 1D topological edge states but also paves the way for further physical investigation on the dynamic behavior of carriers in 1D topological edge states. Moreover, our work indicates the 1D topological edge states have broad application prospects in the field of optoelectronic.

This work is funded by the National Science Foundation of China (NSFC) (11734003), the National Key Research and Development Program of China (2016YFA0300600). Y.G.Y. is supported by NSFC (11574029) and the Strategic Priority Research Program of Chinese Academy of Sciences (XDB30000000).